\documentstyle{article}
\begin{document}
\title{A Simple Algebraic Derivation of the Covariant Anomaly and 
Schwinger Term}
\author{{\bf C.Ekstrand}\\Department of Theoretical Physics, \\Royal Institute of
Technology, \\S-100 44 Stockholm, Sweden}
\date{}
\maketitle

\newcommand{\eq}{\begin{equation}}
\newcommand{\eqend}{\end{equation}}
\newcommand{\eqa}{\begin{eqnarray}}
\newcommand{\eqaend}{\end{eqnarray}}
\newcommand{\nonu}{\nonumber \\ \nopagebreak}
\newcommand{\Ref}[1]{(\ref{#1})}
\newcommand{\A}{{\cal A}}
\newcommand{\G}{{\cal G}}
\newcommand{\F}{{\cal F}}
\newcommand{\U}{{\cal A}}
\newcommand{\cL}{{\cal L}}
\newcommand{\etta}{X}
\newtheorem{theorem}{Theorem}

\begin{abstract}
  An expression for the curvature of the \lq covariant\rq $ $ determinant line bundle is given in even dimensional space-time. The usefulness is guaranteed by its prediction of the covariant anomaly and Schwinger term. It allows a parallel derivation of the consistent anomaly and Schwinger term, and their covariant counterparts, which clarifies the similarities and differences between them. In particular, it becomes clear that in contrary to the case for anomalies, the difference between the consistent and covariant Schwinger term can not be extended to a local form on the space of gauge potentials.
\end{abstract}
\section{Introduction}
 Although the consistent and covariant (chiral) anomaly have many similar features, they also differ in many ways. For example, the consistent anomaly is commonly derived from a variation of the effective action while the covariant anomaly often is determined by the covariant current, defined from the consistent current by addition of a local term, \cite{BZ}. Another example is provided by the descent equations, a set of equations from which the consistent anomaly and Schwinger term not only can be calculated, by their cohomological meaning is also determined. Various attempts, for example \cite{SO,WA,BC}, has been made to create interesting cohomology classes and descent equations for the covariant anomaly (and Schwinger term, \cite{TS}), however, their physical meaning is unclear. 

 Despite these differences, we will show that it is anyway possible to treat the consistent and covariant formalism from a common setting. The advantage of doing so is that the origin of their differences will be illuminated. Further, it allows as simple derivation of the covariant anomaly and Schwinger term as for their consistent counterparts. The results turn out to agree with the ones found in the literature, see for example \cite{BZ,FU,MS} for the covariant anomaly and \cite{HS,NT,K} for the covariant Schwinger term.

 Following Fujikawa \cite{FU}, the covariant anomaly can be obtained from an effective action that has been renormalized covariantly, in contrary to the renormalization leading to the consistent anomaly, see \cite{AG} for instance. This suggests that the covariant anomaly can be obtained from the transgression of the curvature on a \lq covariant\rq $ $ determinant line bundle. We will show that this indeed is possible. The justification of our choice of line bundle is given by the prediction of the covariant Schwinger term as well. It will become clear that the difference between the consistent and covariant anomaly can be extended from gauge orbits to a local form on the space of gauge connections, while this is not true for the difference between the consistent and covariant Schwinger term.

 The paper is organized as follows. To set notations and explain basic ideas, the consistent anomaly and Schwinger term will be derived and considered in section \Ref{sec:CS}. In section \Ref{sec:CO} a parallel treatment will be performed for the covariant anomaly and Schwinger term. The similarities and differences between the consistent and covariant formalism will be discussed in section \Ref{sec:SD}. All these sections deal with Yang-Mills theory and explicit results for lower dimensions will be given in section \Ref{sec:DI}. To explain the modification to other kinds of anomalies and Schwinger terms, the case of diffeomorphisms will also be considered in this section.

\section{Derivation and Interpretation of the Consistent Anomaly and Schwinger term}
\label{sec:CS}
 Let $M$ denote the space-time. 
It is assumed to be a compact oriented $(2n-2)$-dimensional 
Riemannian spin manifold without boundary.
 Consider the principal bundle $P=P(M,G)$, where the gauge group $G$ is assumed to be a compact semi-simple Lie group. For simplicity, only the case when $P$ is trivial will be considered. Gauge potentials are connections on $P$ and can be considered as local one-forms $A$ on $M$ with values in $\mbox{Lie}G$. Gauge transformations are automorphisms on $P$ that projects to the identity map on $M$. They can be considered as local functions $g$ on $M$ taking values in $G$ and obeying suitable gluing properties. 
There is a free action of the group of (base-point preserving) gauge transformation $\G$ on the affine space of gauge potentials $\A$ given by: 
\eq
\label{eq:Ag}
A\mapsto A\cdot g=g^{-1}Ag+g^{-1}dg \quad A\in\A, g\in\G,
\eqend
 where $d$ is the exterior derivative in $M$. This action induces a fibre bundle structure $\pi :\A \rightarrow \A /\G$. Further, to each $A\in\A $ there is associated a Dirac operator $\partial\!\!\! /_A:\Gamma (M,S^+\otimes E)\rightarrow \Gamma (M,S^+\otimes E)$, where $S^{\pm}$ are the positive and negative chirality part of the spin bundle over $M$ and $E$ is an associated bundle to $P$. In local coordinates: 
\eq
\label{eq:DO}
\partial\!\!\! /_A=\sum _{\mu =1}^{2n}\gamma _{\mu }\left( \partial _{\mu }+ \Gamma _{\mu }+ A_{\mu }\right)\left(\frac{1+\gamma _5}{2}\right) .
\eqend
 They satisfy the covariance relation $\partial\!\!\! /_{A\cdot g }=g^{-1}\partial\!\!\! /_Ag $. 

 The effective action can be described as a section of a determinant line bundle $\cL $ over $\A$. Associated with the Quillen metric in $\cL $ is a natural curvature $\F _2$. It has been calculated in ref. \cite{BF2}: 
\eq
\label{eq:O2}
\F _2(\hat{F}^n) =\int _M P\left( \hat{F}^n\right) _{(2n-2,2)}.
\eqend
 The bigrading $(i,j)$ means the part in the expansion of the expression that is an $i$-form on $M$ and a $j$-form on $\A $. $P$ denotes a symmetric invariant polynomial \cite{ZU} (for example, the symmetrized trace in the fibre of $E$) and $\hat{F}=(d+d_{\A })\hat{A}+\hat{A}^2$,
 where $d_{\A }$ is the exterior differential in $\A $ and $\hat{A}(\eta ,\tau )=A(\eta )+a(\tau )(p)$, where $(\eta ,\tau )\in T_pP\times T_A\A$ and $a$ is a connection on $\A\rightarrow\A /\G$. An explicit example is given by $a_A=(D_A^{\ast }D_A)^{-1}D_A^{\ast }$ where $D_A:\mbox{Lie}{\G }\rightarrow T_A\A$ is defined by $D_AX=\left.\frac{d}{dt}\right| _{t=0} A\cdot\exp (tX)$ and $D_A^{\ast }$ is its adjoint, see \cite{AS3}. Since $P\left( \hat{F}^n\right) $ is a closed form on $M\times\A$, it can be written as a coboundary:
\eq
\label{eq:FD}
P\left( \hat{F}^n\right) =  \left( d+d_{\A }\right)\omega _{2n-1}(\hat{A},\hat{F}), 
\eqend
 where 
\eq
\label{eq:QS}
\omega _{2n-1}(\hat{A},\hat{F})=n\int _0^1dt\, P(\hat{A},\hat{F}_t^{n-1}) \quad \hat{F}_t=t\hat{F}+(t^2-t)\hat{A}^2. 
\eqend
 This implies 
\eq
\label{eq:66}
\int _M P\left( \hat{F}^n\right) _{(2n-2,2)}=d_{\A }\int _M\omega _{2n-1}(\hat{A},\hat{F})_{(2n-2,1)}.
\eqend
 Let $i_A(g):=A\cdot g$ and define $\bar{i_A}:P\times {\G }\rightarrow P\times\A$ by $\bar{i_A}=\mbox{id.}_P\times i_A$. The form $\bar{i_A}^{\ast }\hat{A}$ is often denoted by $A+v$, where $v$ is called the ghost. Thus, if supressing the embedding $\bar{i_A}$, we see that the restriction of $\hat{A}$ to vertical vectors in $\A\rightarrow \A /\G$ is equal to $A+v$.
 When restricting to vertical directions and acting on local forms (i.e. polynomials in $A$, $F$, $v$ and $dv$), $d_{\A }$ becomes the BRS operator $\delta $ defined as in \cite{ST}:
\eqa
 \label{eq:BR}
 \delta A & = & -dv -[A,v]\nonu
 \delta v & = & -v^2 \nonu
 \left[\delta ,d\right] & = & 0\nonu
 \delta \, ^2 & = & 0.
\eqaend 
 Graded commutators with respect to the bigrading have been used here and will be used throughout the paper. Using the \lq Russian formula\rq 
\eq 
\left( d+\delta \right)\left( A+v\right) +\left( A+v\right) ^2=F, 
\eqend
the restriction of $\omega _{2n-1}(\hat{A},\hat{F})_{(2n-2,1)}$ to vertical vectors becomes after straight forward computations: 
\eqa
\label{eq:AN}
&& \omega _{2n-2}^1(A,v) := \omega _{2n-1}\left(A+v,\left( d+\delta \right)\left( A+v\right) +\left( A+v\right) ^2\right) _{(2n-2,1)} \nonu 
& = & \omega _{2n-1}( A+v,F)_{(2n-2,1)}\sim 
n(n-1)\int _0^1dt\,\, (1-t)P(dv,A,F_t^{n-2}), 
\eqaend
which is recognized as the (non-integrated) consistent anomaly, see for instance \cite{ZU}. The notation $\sim $ means that we have equality only up to exact forms on $M$ (such forms do not contribute when integrating over $M$). The bigrading that has been used here is with respect to the form degree in $M$ and the ghost degree, respectively.

 Before going into a discussion concerning the meaning of the derivation above, we will explain how the consistent Schwinger term can be obtained. For this, we assume that $M$ has a non-empty boundary $\partial M$, which will be interpreted as the physical space. We will consider the correspondence of eq. \Ref{eq:O2} in this case. The operator $\partial\!\!\! /_A$ will then induce self-adjoint Dirac operators $\partial\!\!\! /_A^{\partial }$ on the $\partial M$ which have to be restricted to certain boundary conditions. This is in analog to the case when considering the index theorem for manifolds with boundary, or the corresponding family index theorem. According to \cite{PZ}, eq. \Ref{eq:O2} becomes now: 
\eq
\label{eq:F2}
\F _2(\hat{F}^n) =\int _M P\left( \hat{F}^n\right) _{(2n-2,2)}+\hat{\eta }.
\eqend
 The $\hat{\eta }$ form accounts for the boundary conditions while the first term is independent of these. It is important to remember that eq. \Ref{eq:F2} does not hold over all of $\A $, but only on a subset $U_{\lambda }=\{ A\in A;\lambda \in \!\!\!\! | \,\,\mbox{spec} \left( \partial\!\!\! /_A^{\partial }\right)\}$. Assume that the $(2n-3)$-dimensional manifold $\partial M$ have the same properties as $M$ and consider the space $\A ^{\partial }$ of gauge potentials on $\partial M$, the corresponding group $\G ^{\partial }$ of gauge transformations, the BRS operator $\delta ^{\partial }$ and the ghost $v^{\partial }$. Let $q:\A ^{\partial }\rightarrow\A$ be a map which extends a gauge potential on the boundary to the whole manifold $M$. The existence of such a map is clear if the structure is of \lq product type\rq $ $ close to $\partial M$, see \cite{EM}. If eq. \Ref{eq:F2} is restricted to $q(U_{\lambda }^{\partial }) $ and pulled-backed by $q$, we obtain
\eq
\label{eq:G2}
q^{\ast }\left(\F _2|_{q(U_{\lambda }^{\partial })}\right) =q^{\ast }\int _M P\left( \hat{F}^n\right) _{(2n-2,2)}+q^{\ast }\left(\hat{\eta }|_{q(U_{\lambda }^{\partial })}\right) , 
\eqend
where $\hat{F}=q(F^{\partial })$ so that $\hat{F}|_{\partial M}=F^{\partial }$.
 From the construction of $\hat{\eta }$ it is clear that $q^{\ast }\left(\hat{\eta }|_{q(U_{\lambda }^{\partial })}\right)$ vanishes when evaluated at vertical vectors in $\A ^{\partial }\rightarrow \A ^{\partial }/\G ^{\partial }$. Thus,  
\eqa
\label{eq:H2}
\left( q^{\ast }\F _2|_{q(U_{\lambda }^{\partial })}\right) _{\mbox{\footnotesize{ver.}}} & = & 
q^{\ast }\left(\int _M\left( \left( d+d_{\A }\right)\omega _{2n-1}(\hat{A},\hat{F})\right) _{(2n-2,2)}\right) _{\mbox{\footnotesize{ver.}}}\nonu 
& = & \int _{\partial M} \omega _{2n-1}(A^{\partial }+v^{\partial },F^{\partial }) _{(2n-3,2)}\nonu 
&&+ q^{\ast }\delta ^{\partial }\int _M\omega _{2n-1}(A+v,F)_{(2n-2,1)}.
\eqaend
 Straight forward computations of the integrand in the first term on the right hand side gives 
\eqa
\label{eq:SN}
&& \omega _{2n-3}^2(A^{\partial },v^{\partial })  :=  
\omega _{2n-1}( A^{\partial }+v^{\partial },F^{\partial })_{(2n-3,2)} \nonu
& \sim & \left\{ \begin{array}{ll}
P(v^{\partial },d^{\partial }v^{\partial }) & n=2 \\
\frac{n(n-1)(n-2)}{2}\int _0^1dt\,\, (1-t)^2P\left(\left( d^{\partial }v^{\partial }\right) ^2,A^{\partial },(F_t^{\partial })^{n-3}\right) & n\geq 3
 \end{array}\right. ,
\eqaend
which is recognized as the (non-integrated) consistent Schwinger term, see for example \cite{ZU}. Also, it is seen that the second term on the right hand side in eq. \Ref{eq:H2} only contributes to the consistent Schwinger term with a coboundary. Thus, the consistent Schwinger term is equal to the curvature of $q^{\ast }\left( \cL _{q(U_{\lambda }^{\partial })}\right)$ evaluated in vertical directions. This agrees with the result of \cite{CMM} where it was shown that $q^{\ast }\left( \cL _{q(U_{\lambda }^{\partial })}\right)$ can be identified with the vacuum line bundle. 

 The rest of this section will be spend on a discussion concerning the cohomological and geometrical meaning of the consistent anomaly and Schwinger term. We start with the anomaly and notice that: $\delta\int _M\omega _{2n-2}^1=0\Leftrightarrow \F _{2,\,\mbox{\footnotesize{ver.}}}=0\Leftrightarrow $ there exists an action of $\G$ on $\cL$ that projects to the action of $\G $ on $\A $. Indeed, these statements are all true since $\F _{2,\,\mbox{\footnotesize{ver.}}}=0$ is a consequence of the Russian formula (this in turn is a consequence of the covariance property satisfied by the Dirac operator in \Ref{eq:DO}). The existence of an action of $\G $ on $\cL $ makes it possible to consider the line bundle $\cL /\G \rightarrow \A /\G $. Many authors prefer to work with $\cL /\G $, however we will restrict to consider $\cL $ since it is only on $\A $ that a parallel treatment can be made for the covariant anomaly. The Wess-Zumino consistency condition $\delta\int _M\omega _{2n-2}^1=0$ implies that $\int _M\omega _{2n-2}^1$ is a representative of an element $[\int _M\omega _{2n-2}^1]_{ {\delta }}\in H^1_{\mbox{\footnotesize{loc.}}}(\mbox{Lie}\G )$, where the cohomology classes $H^k_{\mbox{\footnotesize{loc.}}}(\mbox{Lie}\G )$ are defined by the BRS operator $\delta $ with domain consisting of the $M$-integral of local forms containing $k$ number of ghosts.
 Recall that the consistent anomaly was obtained by writing $\F _2$ as $d_{\A }$ of a 1-form on $\A $ which when restricted to vertical directions could be identified with the $M$-integral of a local form with 1 ghost: $\int _M\omega _{2n-2}^1$. This procedure induces a map:
\eq
\label{eq:FB}
\F _2\mapsto [\int _M\omega _{2n-2}^1]_{ {\delta }}\in H^1_{\mbox{\footnotesize{loc.}}}(\mbox{Lie}\G )
\eqend
which maps the curvature of $\cL $ to the cohomology class of the consistent anomaly. We will refer to this map as the transgression map. By abuse of language we will sometimes also call $ \F _2\mapsto \int _M\omega _{2n-2}^1$ the transgression map, keeping in mind that it is only well defined up to coboundaries. One advantage of working on $\A /\G $ is now clear: Since $\F _{2,\,\mbox{\footnotesize{ver.}}}=0$, the push-forward $\pi _{\ast }\F _2$ can be constructed. It is the curvature on $\cL /\G $. In terms of it, transgression induces a homomorphism from $H^2(\A /\G )$ to $H^1_{\mbox{\footnotesize{loc.}}}(\mbox{Lie}\G )$ according to 
\eq
[\pi _{\ast }\F _2]\rightarrow [\int _M\omega _{2n-2}^1]_{ {\delta }}.
\eqend
 As we will see in the next section, a corresponding homomorphism does not exist for the covariant anomaly. 

 There is a bijective correspondence between connections $\nabla $ on $\cL $ and connection 1-forms $A _{\cL ^{\times }}$ on $\cL ^{\times }$ (=$\cL \setminus s_0$, where $s_0$ is the zero section of $\cL$). It is given by the claim that $\nabla s = s^{\ast }A _{\cL ^{\times }}\otimes s$ should hold for any locally defined section $s$ of $\cL ^{\times }$. From this it is seen that the curvature $\F _2$ locally can be written as $d_{\A }\left(\left(\nabla s^{\prime }\right) /s^{\prime }\right) $, where $s^{\prime }$ is a locally defined non-vanishing section of the line bundle $\cL $ with connection $\nabla $. Eq. \Ref{eq:66} then implies that $\int _M\omega _{2n-1}(\hat{A},\hat{F})=\nabla s^{\prime }/s^{\prime }+d_{\A }\beta$, for some function $\beta $ on $\A $. A change of section $s=s^{\prime }\exp (\beta )$ gives: 
\eq
\label{eq:ON}
\int _M\omega _{2n-1}(\hat{A},\hat{F})=\frac{\nabla s}{s}.
\eqend
 Thus, the consistent anomaly is given by the 1-form $(\nabla _{\mbox{\footnotesize{ver.}}}s)/s$ identified with a local form with 1 ghost. Notice that a change of section $s^{\prime \prime}=s\exp (\beta _{\mbox{\footnotesize{loc.}}})$, where $\beta _{\mbox{\footnotesize{loc.}}}$ is a a local form with no ghosts, gives a change in the consistent anomaly by a coboundary $\delta\beta _{\mbox{\footnotesize{loc.}}}$ and leaves therefore the cohomology class of the consistent anomaly unchanged. Conversely, it is clear that for every change in the representative of the cohomology class, it is possible to multiply $s$ with the exponent of a local form so that \Ref{eq:ON} still holds. A section $s$ is said to be $\mbox{Lie}\G $-equivariant if $\nabla _{\mbox{\footnotesize{ver.,}} X_A}s=s$, $\forall X\in \mbox{Lie}\G $, where $X_A=\frac{d}{dt}|_{t=0}A\cdot \exp (tX)$. Thus, the cohomology class of the consistent anomaly is the obstruction in $\mbox{Lie}\G$-equivariance for a certain locally defined non-vanishing section $s$ of the line bundle $\cL$, with connection $\nabla $. 

 A corresponding analysis for the consistent Schwinger term will now be performed. It can easily be checked that the cocycle relation $\delta ^{\partial }\int _{\partial M}\omega _{2n-3}^2=0$ is fulfilled. This is a consequence of the Russian formula. Thus, the consistent Schwinger term defines an element in $H^2_{\mbox{\footnotesize{loc.}}}(\mbox{Lie}\G ^{\partial })$. It is easy to generalize the transgression map, defined in \Ref{eq:FB}, to a map 
\eq
\label{eq:FC}
\F _3\mapsto [\int _{\partial M}\omega _{2n-3}^2]_{ {\delta ^{\partial }}}\in H^2_{\mbox{\footnotesize{loc.}}}(\mbox{Lie}\G ^{\partial })
\eqend
 with 
\eq
\label{eq:NU}
\F _3=\int _{\partial M} P\left( (\hat{F}^{\partial })^n\right) _{(2n-3,3)}=
d_{\A ^{\partial }}\int _{\partial M}\omega (\hat{A},\hat{F})_{(2n-3,2)}.
\eqend
 Similar to the case for the consistent anomaly, this induces a homomorphism from $H^3(\A ^{\partial }/\G ^{\partial })$ to $H^2_{\mbox{\footnotesize{loc.}}}(\mbox{Lie}\G ^{\partial })$ that maps $[\pi _{\ast }\F _3]$ to the right hand side of \Ref{eq:FC}.
 Observe that the cocycle relation is equivalent with $\F _{3,\,\mbox{\footnotesize{ver.}}}=0$. 

The consistent Schwinger term was obtained by writing $\F _2$ as a space-time (without boundary) integral of a form and considering the corresponding expression when space-time has a boundary. This gives an expression containing two terms, see eq. \Ref{eq:F2}. The first term is the one obtained if we naively let space-time have a boundary in the formula for $\F _2$. The second term accounts for the boundary conditions that the induced Dirac operators on the boundary have to fulfill. Since the integrand in the first term only depends on the intrinsic properties of $M$, we will refer to this term as the \lq intrinsic part\rq . Thus, the consistent Schwinger term is obtained by replacing the boundary-less space-time in the formula for $\F _2$ with a space-time with boundary, keeping only the intrinsic part, pull-back by $q$ and restricting to vertical vectors. Remember that $\F _2$ is the curvature on $\cL $, the line bundle on which the effective action is described by a section. Since the second term in \Ref{eq:G2} vanishes when restricted to vertical vectors, the consistent Schwinger term is in fact equal to the curvature of a locally defined line bundle over $\A ^{\partial }$. However, this interpretation will not be used in this paper since there is no correspondence of it for the covariant Schwinger term, as will be explained in the next section.

\section{Derivation and Interpretation of the Covariant Anomaly and Schwinger term}
\label{sec:CO}
 In the previous section we reviewed the result that the non-integrated consistent anomaly and Schwinger term are given by the parts in the expansion of 
\eq
\label{eq:ref}
\omega _{2n-1}\left(A+v,\left( d+\delta \right) \left( A+v\right) +\left( A+v\right) ^2\right)
\eqend
 that is linear and quadratic in the ghost, respectively. To be consistent with earlier notation we should of course replace $A$ with $A^{\partial }$, $d$ with $d^{\partial }$, $v$ with $v^{\partial }$ and $\delta $ with $\delta ^{\partial }$, for the consistent Schwinger term. A close inspection of the consistent anomaly and Schwinger term reveals that the breakdown of gauge invariance depends on the appearance of the BRS operator $\delta $ in this expression. This makes it plausible that the covariant anomaly and Schwinger term can be derived by naively dropping $\delta $. These obstructions have been obtained in many different ways, see for example \cite{BZ,FU} for the covariant anomaly and \cite{TS,HS,NT,K}
 for the covariant Schwinger term, and they all agree with the result
\eq
\label{eq:AVP}
\tilde{\omega }_{2n-2}^1(A,v) \sim nP(v, F^{n-1}) 
\eqend
for the covariant anomaly and 
\eq
\label{eq:QQP}
\tilde{\omega }_{2n-3}^2(A^{\partial },v^{\partial }) \sim  \frac{n(n-1)}{2}P(v^{\partial },-\delta ^{\partial }A^{\partial }, (F^{\partial })^{n-2})
\eqend
for the covariant Schwinger term.
\begin{theorem}
 The non-integrated covariant anomaly and Schwinger term are given by the parts in the expansion of 
\eq
\label{eq:21}
\omega _{2n-1}\left(A+v,d\left( A+v\right) +\left( A+v\right) ^2\right) 
\eqend
that is linear respective quadratic in the ghost. (To have consistent notation with the rest of the paper, replace $A$ with $A^{\partial }$, and so on, for the covariant Schwinger term)
\end{theorem}
 The theorem can be proven by explicit computations and comparison with \Ref{eq:AVP} and \Ref{eq:QQP}. Since the computations for the covariant Schwinger term are the most difficult ones, we present them in the Appendix. Observe that the fact that the covariant Schwinger term is the part of \Ref{eq:21} that is quadratic in the ghost is not in agreement with the corresponding erroneous formula obtained earlier in the literature, for example in \cite{TS}. 

 The fact that theorem 1 is true implies that the covariant anomaly can be obtained from a line bundle $\tilde{\cL }$ defined as $\cL $, but with curvature $\tilde{\F }_2$ such that the covariant correspondence of $\omega (\hat{A}, \hat{F})$ in eq. \Ref{eq:FD} is a form that restricts to $\omega _{2n-1}\left(A+v,d\left( A+v\right) +\left( A+v\right) ^2\right) $ at vertical directions. It is easy to see that such a construction is possible, for example, $\A $ affine implies that the integral of $\tilde{\F }_2$ over a subset of $\A $ without boundary is zero, which guarantees the existence of a line bundle $\tilde{\cL }$ with curvature $\tilde{\F }_2$. An explicit example is given by 
\eq
\tilde{\F }_2=d_{\A }\int _M \omega _{2n-1}(\hat{A}, d\hat{A}+\hat{A}^2)_{(2n-2,1)}.
\eqend

 We define the cohomology class of the covariant anomaly $[\int _M\tilde{\omega }_{2n-2}^1]$ to be $\int _M\tilde{\omega }_{2n-2}^1$ modulo $\delta \omega $, where $\omega $ is a local form. Then, transgression gives the map
\eq
\label{eq:3B}
\tilde{\F }_2\mapsto [\int _M\tilde{\omega }_{2n-2}^1]
\eqend
which maps the curvature of $\tilde{\cL }$ to the covariant anomaly. Consider the following equivalent relations: $\delta\int _M\tilde{\omega }_{2n-2}^1\neq 0\Leftrightarrow \tilde{\F }_{2,\,\mbox{\footnotesize{ver.}}}\neq 0\Leftrightarrow $ there does not exists an action of $\G$ on $\tilde{\cL }$ that projects to the action of $\G $ on $\A $. That they are true follows from the fact that $d\delta \tilde{\omega }_{2n-2}^1$ is not identical to zero, as is easy to check. This means that $[\int _M\tilde{\omega }_{2n-2}^1]$, in contrary to $[\int _M\omega _{2n-2}^1]$, does not define any cohomology class $[...]_{ {\delta }}$. Also, it implies that it is not possible to work on $\A /\G $ for the covariant anomaly. However, it is easy to see that the cohomology class of the covariant anomaly can be interpreted geometrically in a corresponding way to what was done for the consistent anomaly. In fact, the considerations in the previous section goes through word by word if putting a tilde on all objects that appears. Thus, the cohomology class of the covariant anomaly is the obstruction in $\mbox{Lie}\G $-equivariance for a certain local non-vanishing section $\tilde{s}$ of $\tilde{\cL }$, with connection $\tilde{\nabla }$.

 Theorem 1 also implies that the covariant Schwinger term can be obtained in a similar way as the consistent Schwinger term was derived: First, write $\tilde{\F }_2$ as the integral over $M$ of a form, a covariant correspondence of eq. \Ref{eq:O2}. Then, consider the expression for $\tilde{\F }_2$ obtained by naively replacing $M$ with a manifold with boundary in eq. \Ref{eq:O2}. After pull-back by $q$ and restriction to vertical directions, the covariant Schwinger term is obtained. Notice that when these steps were performed for the consistent case we saw that the Schwinger term was equal to a curvature along gauge directions for a line bundle. This comes form eq. \Ref{eq:G2} and the fact that the $q$ pull-back of $\hat{\eta }$ vanishes at gauge directions. In the covariant case, there is no direct correspondence of the family index theorem and eq. \Ref{eq:G2}. Especially, the covariant Schwinger term can not be interpreted as the curvature of a line bundle (if it could, it would satisfy the cocycle relation, which is not true).

 Transgression gives the map: 
\eq
\label{eq:4B}
\tilde{\F }_3\mapsto [\int _{\partial M}\tilde{\omega }_{2n-3}^2] ,
\eqend
where $\tilde{\F }_3$ is given by 
\eq
\tilde{\F }_3=d_{\A ^{\partial }}\int _{\partial M} \omega _{2n-1}\left( \hat{A}^{\partial }, d^{\partial }\hat{A}^{\partial }+\left( \hat{A}^{\partial }\right) ^2\right)_{(2n-3,2)},
\eqend
for example.
The notation $[...]$ is the equivalence relation defined so that two representatives are equal if and only if they differ by $\delta ^{\partial }\omega $, for some local form $\omega $. It is easy to check that $d^{\partial }\delta ^{\partial }\tilde{\omega }_{2n-3}^2$ is not identical to zero for $n\geq 3$. It implies that the covariant Schwinger term $[\int _{\partial M}\tilde{\omega }_{2n-3}^2]$, $n\geq 3$, does not obey the cocycle relation and does therefore not define any cohomology class with respect to $\delta ^{\partial }$. This fact is equivalent with that $\tilde{\F }_{3,\,\mbox{\footnotesize{ver.}}}\neq 0$, $n\geq 3$. This means that it is not possible to work on $\A /\G $ here either. Notice that $n=2$ is an exception where $d^{\partial }\delta ^{\partial }\tilde{\omega }_{1}^2=0$ and $\delta ^{\partial }\int _{\partial M}\tilde{\omega }_{1}^2=0 \Leftrightarrow \tilde{\F }_{3,\,\mbox{\footnotesize{ver.}}}= 0$ are true.

 There is a natural normalization factor for the consistent anomaly and Schwinger given by the claim that $\pi _{\ast }\F _2$ and $\pi _{\ast }\F _3$ should obey the so-called integrality condition. That means that the integral of $\pi _{\ast }\F _2$ over any closed surface in $\A /\G$ is equal to an integer (or sometimes, an integer times a constant factor), and corresponding for $\pi _{\ast }\F _3$ and a closed volume in $\A /\G$. Since it is not possible to work on $\A /\G$ for the covariant anomaly and Schwinger term (for $n\geq 3$) it is not as easy to find the correct normalization factor in this case. We choose to define the normalization factor for the covariant anomaly and Schwinger term simply by dropping the BRS operator in the formula \Ref{eq:ref} for their consistent counterparts, normalized by the integrality condition.

 Fujikawa \cite{FU} showed that the covariant anomaly can be obtained as the obstruction in $\G $-equivariance for some determinant \lq function\rq . The determinant \lq function\rq $ $ he used was a gauge covariant renormalized effective action. A correct mathematical interpretation would be to view Fujikawa's determinant as a section of a line bundle that to some perspective is covariant. Since we obtain the same result as Fujikawa, this motivates us to use the name \lq covariant\rq $ $ determinant line bundle for $\tilde{\cL }$. As mentioned in the introduction, the final justification for the name is that it predicts the covariant Schwinger term as well. That is, we get the same result as in \cite{HS,NT,K} where the covariant Schwinger term was obtained from a renormalization of the commutator of covariant \lq Gauss law operators\rq . With the latter we mean the ordinary (consistent) Gauss law operators with the consistent current replaced with the covariant current.

\section{A comparison of the consistent and covariant formalism}
\label{sec:SD}
 In the consistent formalism, the Russian formula implies 
\eqa
\label{eq:17}
 (d+\delta )\omega _{2n-1}(A+v,F) & = & P
\left(\left(\left( d+\delta\right)\left( A+v\right)+\left( A+v\right) ^2\right) ^n\right)\nonu 
& = & P(F^n) =d\omega _{2n-1}(A,F).
\eqaend
 Equating forms with the same bigrading gives the descent equations:
\eqa
P(F^n) & = & d\omega _{2n-1}(A,v)\nonu
\delta  \omega _{2n-1}(A,v) & = & -d\omega _{2n-2}^1(A,v)\nonu
\delta \omega _{2n-2}^1(A,v) & = & -d\omega _{2n-3}^2(A,v)\nonu
& ... &\nonu
\delta  \omega _{0}^{2n-1}(A,v) & = & 0,
\eqaend
where $\omega _{2n-1-k}^k(A,v) := \omega _{2n-1}\left( A+v,F\right) _{(2n-1-k,k)}$, see \cite{ZU}. This provides an alternative way to calculate the consistent anomaly and Schwinger term (for the latter it must of course be assumed that the operators and variables belong to the physical space, which in this paper means that the index $\partial $ should be used). Further, their cohomological meaning is also determined by this set of equations. 

 For the covariant anomaly and Schwinger term, there is no similar set of equations. For example, eq. \Ref{eq:17} is based on the fact that $\F _{2,\,\mbox{\footnotesize{ver.}}}=0$, a relation whose covariant correspondence is not true. In the previous section we showed that this implies that it is not possible to assign a cohomological meaning to the covariant anomaly and Schwinger term in the same way as was made for their consistent counterparts. However, there has been attempts to construct interesting \lq covariant\rq $ $ descent equations, see for example \cite{SO,WA,TS}. 

 After having mentioned this important difference between the consistent and covariant formalism, we will now consider how the two formalisms can be related. First of all, we would like to point out that they can be interpreted in a similar way. Both the consistent and covariant anomaly are the obstruction in $\G $-equivariance for sections of line bundles. The consistent and covariant Schwinger term are both obtained by replacing $M$ with a manifold with boundary in the formula for the effective action, pull-back by $q$ and restrict to vertical directions.

 An informative way to study the relation between the consistent and covariant anomaly is to consider their difference: 
\eqa
\label{eq:24}
\omega _{\mbox{\footnotesize{loc.}},2n-2}^1(A,v) & := & \omega _{2n-2}^1(A,v)-\hat{\omega }_{2n-2}^1(A,v)\nonu
& = & \omega _{2n-1}\left(A,\left( d+\delta \right) A +A^2\right) _{(2n-2,1)}\nonu
& \sim & n(n-1)\int _0^1dt\,\, tP(A,-\delta A,F_t^{n-2}).
\eqaend
 It is interesting to note that this form could have been obtained by transgression of the curvature ${\F }_{\mbox{\footnotesize{loc.}},2}$ on the line bundle $\cL _{\mbox{\footnotesize{loc.}}}$ defined in the same way as $\cL $, but with $\hat{A}$ replaced with $A$ (this is the universal connection considered in \cite{FA} for instance).
 $\cL _{\mbox{\footnotesize{loc.}}}$ differs from $\cL $ and $\tilde{ \cL }$ in an essential aspect which now will be explained. A form on $\A$ will be referred to as being $\A $-local if it can be written as a polynomial in $A$, $dA$ and $d_{\A }A$. An example is given by $\omega _{2n-1}(A,(d+d_{\A })A+A^2)$. Its restriction to vertical directions in $\A $ is equal to the form $\omega _{2n-1}(A,(d+\delta)A+A^2)$ that appears in eq. \Ref{eq:24}. Thus, the form 
$\omega _{2n-1}(A,(d+\delta)A+A^2)$, related to $\cL _{\mbox{\footnotesize{loc.}}}$, can be continuously extended to an $\A $-local form. This is however not possible for $\omega _{2n-1}(A+v,F)$ and $\omega _{2n-1}(A+v,d(A+v)+(A+v)^2)$, related to $\cL $ and $\tilde{ \cL }$. That such an extension exists for $\omega _{\mbox{\footnotesize{loc.}},2n-2}^1(A,v)$ can be seen from the fact that the right hand side of eq. \Ref{eq:24} contains the ghost only in the combination $\delta A$, while it is not so for eq. \Ref{eq:AN} and \Ref{eq:AVP}. Thus, the difference between the consistent and covariant anomaly can be continuously extended to an $\A $-local form, \cite{STORA}. It is now clear that the covariant anomaly can be obtained from the transgression of the curvature $\F _2-{\F }_{\mbox{\footnotesize{loc.}},2}$ on the line bundle $\cL \otimes ({\cL }\mbox{\footnotesize{loc.}})^{\ast }$. This fact is equivalent with the approach of obtaining the covariant anomaly with the introduction of a background connection (coming from $\cL _{\mbox{\footnotesize{loc.}}}$) which has been performed in the literature \cite{MS,K}. This suggests the possibility of that also the covariant Schwinger term can be obtained from a background connection, or equivalently from $\cL \otimes ({\cL }\mbox{\footnotesize{loc.}})^{\ast }$. However, this is not true since we showed in the previous section that the covariant Schwinger term is obtained from $\tilde{\cL }$. In fact, if it would be possible to continuously extend $\omega _{2n-3}^2-\tilde{\omega }_{2n-3}^2$ to an $\A ^{\partial }$-local form, then so would also be the case for $\omega _{2n-3}^2-\tilde{\omega }_{2n-3}^2-\omega _{\mbox{\footnotesize{loc.}},2n-3}^2$ and $\delta ^{\partial }\left(\omega _{2n-3}^2-\tilde{\omega }_{2n-3}^2-\omega _{\mbox{\footnotesize{loc.}},2n-3}^2\right)$, where $\omega _{\mbox{\footnotesize{loc.}},2n-3}^2$ is defined by
\eq
\omega _{\mbox{\footnotesize{loc.}},2n-3}^2(A^{\partial },v^{\partial }):=\omega _{2n-1}(A^{\partial },(d^{\partial }+\delta ^{\partial })A^{\partial }+(A^{\partial })^2)_{(2n-3,2)}.
\eqend
 By similar computation techniques as in the Appendix, it follows that 
\eqa
&&\omega _{2n-3}^2-\tilde{\omega }_{2n-3}^2-\omega _{\mbox{\footnotesize{loc.}},2n-3}^2(A^{\partial },v^{\partial })=n(n-1)P\left( v^{\partial },\delta ^{\partial }A^{\partial }, (F^{\partial })^{n-2 }\right) \nonu
&&+\delta ^{\partial }n(n-1)\int _0^1dt\, tP\left( v^{\partial },A^{\partial }, (F_t^{\partial })^{n-2 }\right) ,\quad n\geq 3,
\eqaend
from which it is clear that $\delta ^{\partial }\left(\omega _{2n-3}^2-\tilde{\omega }_{2n-3}^2-\omega _{\mbox{\footnotesize{loc.}},2n-3}^2\right)$, $n\geq 3$, can not be continuously extended to an $\A ^{\partial }$-local form. For $n=2$ this fact follows directly from a dimensional reasoning.
 Thus, the difference between the consistent and covariant Schwinger term can not be continuously extended to an $\A ^{\partial }$-local form, in contrary to the case for the anomaly on $\A $. Notice however that for $n=2$ there exist a covariant representative of the cohomology class of the consistent Schwinger term. It differs with a sign from our definition of the covariant Schwinger term.

\section{Low Dimensional Covariant Schwinger Terms for Yang-Mills and Diffeomorphisms}
\label{sec:DI}
 We here write down explicit expressions for the covariant Yang-Mills and Diffeomorphism Schwinger term in 1 and 3 dimensional space. For simplicity, we will start with the case of Yang-Mills. Assume thus that $P(\cdot )$ is the symmetrized trace times the factor $c=4\pi(i/2\pi )^n/n!$ and consider eq. \Ref{eq:QQP} for $n=2$ and $n=3$:
\eqa
\int _{\partial M}\tilde{\omega }_{1}^2(A^{\partial },v^{\partial }) & = & 
 \frac{1}{2\pi }\int _{\partial M}\mbox{tr}(v^{\partial }\delta ^{\partial }A^{\partial }) \nonu
\int _{\partial M}\tilde{\omega }_{3}^2(A^{\partial },v^{\partial }) & = & 
 \frac{i}{8 \pi ^2}\int _{\partial M}\mbox{tr}\left( \left(v^{\partial }F^{\partial }+F^{\partial }v^{\partial }\right)\delta ^{\partial }A^{\partial }
\right) .
\eqaend
 
 We will now consider these expressions when $\partial M$ is an Euclidean space and all forms and functions have compact support therein. Evaluated on infinitesimal gauge transformations $X ^{\partial }Y^{\partial }\in \mbox{Lie}\G ^{\partial }$ it gives:
\eqa
&&\int _{\bf R}\tilde{\omega }_{1}^2(A^{\partial },v^{\partial }) (X^{\partial },Y^{\partial })  =  \frac{1}{2\pi }\int _{\bf R}\mbox{tr}\left(  X^{\partial }D_{A^\partial }Y^{\partial }\right) dx \nonu
&&\int_{{\bf R}^3}\tilde{\omega }_{3}^2(A^{\partial },v^{\partial })(X^{\partial },Y^{\partial }) \nonu
 & = & \frac{i}{8 \pi ^2}\int _{{\bf R}^3}\mbox{tr}\left( \left( X^{\partial }D_{A_k^\partial }Y^{\partial }-(D_{A_k^\partial }Y^{\partial })X^{\partial }\right) F^{\partial }_{ij}\epsilon _{ijk}\right) d^3x,
\eqaend
where $A^{\partial }=A^{\partial }_xdx$, $A^{\partial }=\sum _{k=1}^3A^{\partial }_kdx^k$, $F^{\partial }=\sum _{i,j=1}^3F^{\partial }_{i,j}dx^idx^j$ and $\epsilon _{ijk}$ is the anti-symmetric tensor with $\epsilon _{123}=1$. This can be compared with the corresponding formulas for the consistent case:
\eqa
&&\int _{\bf R}\omega _{1}^2(A^{\partial },v^{\partial }) (X^{\partial },Y^{\partial })  =  -\frac{1}{2\pi }\int _{\bf R}\mbox{tr}\left( X^{\partial }\partial _xY^{\partial }\right) dx \nonu
&&\int _{{\bf R}^3}\omega _{3}^2(A^{\partial },v^{\partial })(X^{\partial },Y^{\partial }) \nonu
& = & -\frac{i}{24 \pi ^2}\int _{{\bf R}^3}\mbox{tr}\left( \left( \left( \partial _iX^{\partial }\right) \partial _jY^{\partial }-\left(\partial _i Y^{\partial }\right)\partial _jX^{\partial }\right) A^{\partial }_k\epsilon _{ijk}\right) d^3x.
\eqaend

 It is known \cite{BZ,GN} that the structure of the diffeomorphism anomaly is similar to the gauge anomaly. The same is true also for the Schwinger term, \cite{EM}. The only thing that differs is the interpretation of the objects in the formulas. For example, the space $\A $ of gauge connections $A$ should be replaced by the space of Christoffel connections $\Gamma $. For a vector $\xi $ generating a diffeomorphism, the ghost is locally given by $v_{\mu }^{\lambda }=\partial _{\mu }\xi ^{\lambda }$. Thus, also in this case, the formulas for the consistent and covariant anomaly and Schwinger term are given by eq. \Ref{eq:AN}, \Ref{eq:SN}, \Ref{eq:AVP} and \Ref{eq:QQP}. We will make this explicit by calculating the covariant Schwinger term in 1, 3 and 5 dimensional space. The relevant formula to use is eq. \Ref{eq:QQP} with $P(\cdot )=4\pi\hat{A}(\cdot )$, where $\hat{A}(\cdot )$ is the $A$-roof function appearing in the index theorem. It is not to be confused with the potential $\hat{A}$ considered before. It is defined by 
\eqa
\hat{A}(R^{\partial }) & = & \mbox{det}^{1/2}\left( \frac{iR^{\partial }/4\pi }{\sinh (iR^{\partial }/4\pi)}\right) =
1+\left(\frac{1}{4\pi }\right) ^2\frac{1}{12}\mbox{tr}\left( R^{\partial }\right) ^2 \nonu
&&+ \left(\frac{1}{4\pi }\right) ^4\left( \frac{1}{288}\left( \mbox{tr}\left( R^{\partial }\right) ^2\right) ^2 + \frac{1}{360}\mbox{tr}\left( R^{\partial }\right) ^4\right) + ...\quad ,
\eqaend
where $R^{\partial }$ is the Riemannian curvature on $\partial M$, see for instance\cite{AG}.
 For $n=2$ we obtain
\eq
\int _{\partial M}\tilde{\omega }_1^2(\Gamma ^{\partial },v^{\partial })=-\int _{\partial M}P(v^{\partial },\delta ^{\partial }\Gamma ^{\partial })=
-\frac{1}{4\pi }\frac{1}{12}\int _{\partial M}\mbox{tr}\left(
v^{\partial }\delta ^{\partial }\Gamma ^{\partial }\right) .
\eqend
 For n=3:
\eq
\int _{\partial M}\tilde{\omega }_3^2(\Gamma ^{\partial },v^{\partial })=0
\eqend
and for n=4:
\eqa
&&\int _{\partial M}\tilde{\omega }_5^2(\Gamma ^{\partial },v^{\partial })  = \int _{\partial M} P\left( v^{\partial },\delta ^{\partial }\Gamma ^{\partial }, (R^{\partial })^2\right) \nonu
& = &
-\left(\frac{1}{4\pi }\right) ^3\frac{1}{576}\int _{\partial M}\left(\mbox{tr}\left( v^{\partial }\delta ^{\partial }\Gamma ^{\partial }\right) \mbox{tr}\left( (R^{\partial })^2\right)+\mbox{tr}\left( v^{\partial }R^{\partial }\right)\mbox{tr}\left( R^{\partial }\delta ^{\partial }\Gamma ^{\partial }\right)\right) 
\nonu
&&-\left(\frac{1}{4\pi }\right) ^3\frac{1}{3240}\int _{\partial M}\mbox{tr}\left( v^{\partial }(R^{\partial })^2\delta ^{\partial }\Gamma ^{\partial }+v^{\partial }R^{\partial }(\delta ^{\partial }\Gamma ^{\partial })R^{\partial }+v^{\partial }(\delta ^{\partial }\Gamma ^{\partial })(R^{\partial })^2\right) .\nonu
\eqaend
\thanks{\bf Acknowledgments:}
 I thank J.Mickelsson for drawing my attention to covariant Schwinger terms and for his remarks on a preliminary version of the paper.  
\newpage
\section*{Appendix: Calculating the Covariant Schwinger Term}
\label{sec:SC}
 We will here give the details of the calculations needed to prove Theorem 1 for the case of the covariant Schwinger term. Eq. \Ref{eq:QS} gives 
\eqa
&&\omega _{2n-1}\left(A^{\partial }+v^{\partial },d^{\partial }\left( A^{\partial }+v^{\partial }\right) +\left( A^{\partial }+v^{\partial }\right) ^2\right) _{(2n-3,2)} \nonu
& = & n\int _0^1dt\, P\left( A^{\partial }+v^{\partial },\left(td^{\partial }(A^{\partial }+v^{\partial })+t^2(A^{\partial }+v^{\partial })^2\right) ^{n-1}\right) _{(2n-3,2)}\nonu 
& = & \frac{n(n-1)}{2}\int _0^1dt\, \Big( 2P\left( v^{\partial }, tD^{\partial }_tv^{\partial }, (F^{\partial }_t)^{n-2}\right) \nonu
&& +2P\left( A^{\partial }, t^2(v^{\partial })^2, (F_t^{\partial })^{n-2}\right) + (n-2)P\left( A^{\partial }, tD^{\partial }_tv^{\partial }, tD^{\partial }_tv^{\partial }, (F^{\partial }_t)^{n-3}\right)\Big) , \nonu
\eqaend
where $D^{\partial }_t:=d^{\partial }+[A^{\partial },\cdot \, ]$ and the commutator is graded. Using $D^{\partial }_tF^{\partial }_t=0$ and making a \lq partial integration\rq $ $ with respect to the operator $D^{\partial }_t$ leads to the following relation for the last term:
\eqa
&& (n-2)P\left( A^{\partial }, tD^{\partial }_tv^{\partial }, tD^{\partial }_tv^{\partial }, (F^{\partial }_t)^{n-3}\right) \nonu
& \sim & (n-2)P\left( tv^{\partial }, tD^{\partial }_tv^{\partial }, D^{\partial }_tA^{\partial }, (F^{\partial }_t)^{n-3}\right) \nonu 
&& -(n-2)P\left( A^{\partial }, t^2(D^{\partial }_t)^2v^{\partial },v^{\partial }, (F^{\partial }_t)^{n-3}\right) .
\eqaend
 The identities $(D^{\partial }_t)^2v^{\partial }=[F^{\partial }_t,v^{\partial }]$ and $\frac{d}{dt}D^{\partial }_tv^{\partial }=[A^{\partial },v^{\partial }]$ applied to the last term in this expression gives:
 \eqa
(n-2)P\left( A^{\partial }, t^2(D^{\partial }_t)^2v^{\partial },v^{\partial }, (F_t^{\partial })^{n-3}\right) & = & 2P\left( A^{\partial }, t^2(v^{\partial })^2, (F_t^{\partial })^{n-2}\right) \nonu
&& -P\left( \frac{d}{dt}D^{\partial }_tv^{\partial }, t^2v^{\partial }, (F^{\partial }_t)^{n-2}\right) . \nonu
\eqaend
 When putting it all together we obtain the final result:
\eqa
&&\omega _{2n-1}\left(A^{\partial }+v^{\partial },d^{\partial }\left( A^{\partial }+v^{\partial }\right) +\left( A^{\partial }+v^{\partial }\right) ^2\right) _{(2n-3,2)} \nonu
& \sim & 
\frac{n(n-1)}{2}\int _0^1dt\, \Bigg( 2P\left( v^{\partial }, tD^{\partial }_tv^{\partial }, (F^{\partial }_t)^{n-2}\right) \nonu
&& +
P\left( \frac{d}{dt}D^{\partial }_tv^{\partial }, t^2v^{\partial }, (F^{\partial }_t)^{n-2}\right) +(n-2)P\left( tv^{\partial }, tD^{\partial }_tv^{\partial }, D^{\partial }_tA^{\partial }, (F^{\partial }_t)^{n-3}\right) \Bigg) \nonu
& = &
\frac{n(n-1)}{2}\int _0^1dt\, \frac{d}{dt}P\left( tv^{\partial }, tD^{\partial }_tv^{\partial }, (F^{\partial }_t)^{n-2}\right) \nonu
& = & \frac{n(n-1)}{2}P\left( v^{\partial }, -\delta ^{\partial }A^{\partial }, (F^{\partial })^{n-2}\right) .
\eqaend 

\end{document}